\documentclass{article}
\PassOptionsToPackage{numbers,sort&compress}{natbib}
\usepackage[preprint]{neurips_2026}
\usepackage[T1]{fontenc}
\usepackage{tgtermes}
\usepackage{tgheros}

\usepackage{graphicx,array,siunitx}
\usepackage{xcolor}
\usepackage{amsmath}
\usepackage[colorlinks=true, linkcolor=blue, citecolor=blue]{hyperref}

\newcommand{\kjmol}{\mathrm{kJ\,mol^{-1}}}
\newcommand{\kcalmol}{\mathrm{kcal\,mol^{-1}}}
\newcommand{\angstrom}{\mathring{\mathrm{A}}}

\title{\texttt{PLUMED-JAX}: GPU-native  enhanced sampling for atomistic simulations}
\author{%
  Venkat Kapil \\
  Department of Physics and Astronomy, University College London \\
  7-19 Gordon St, London WC1H 0AH, UK \\
  Thomas Young Centre and London Centre for Nanotechnology \\
  9 Gordon St, London WC1H 0AH, UK \\
  \texttt{v.kapil@ucl.ac.uk}
}

\begin{document}
\maketitle

\begin{abstract}
We present \texttt{PLUMED-JAX}, an AI-assisted adaptation of selected \texttt{PLUMED} collective variables, umbrella sampling, and metadynamics for \texttt{JAX}-based atomistic simulations, developed using \texttt{gpt-6.1-sol} with reasoning effort set to \texttt{high} within the \texttt{Codex} harness.
Integration with \texttt{JAX-MD} supports compiled molecular dynamics, batched replicas, walkers sharing a bias, and execution across multiple graphics processing units.
We validate the framework against \texttt{PLUMED} and exact free energy benchmarks, and demonstrate applications to butane, silicon crystallization, and lithium transport, with severalfold improvements in computational throughput for both classical forcefields and machine learning interatomic potentials.
\end{abstract}

\section{Introduction}
\noindent Enhanced sampling methods enable the exploration of rare atomistic phenomena and their underlying free energy landscapes~\cite{tuckerman_statistical_2010}.
Recent applications illustrate the widespread impact of enhanced sampling such as across zeolite catalysis~\cite{bocus_confined_2025}, ribosomal substrate selection~\cite{watson_atomistic_2023}, phase change materials for heat storage~\cite{liu_layer-by-layer_2024}, and the physics of nanoconfined water~\cite{kapil_first-principles_2022}.
In this context, the \texttt{PLUMED} library ~\cite{bonomi_plumed_2009,tribello_plumed_2014,bonomi_promoting_2019,rapetti_making_2026} has played a central role in making enhanced sampling accessible and reproducible across forcefield and electronic structure codes. 
Notably, its collective variable library and modular design facilitate the development and testing of new sampling methods~\cite{pfaendtner_efficient_2015,valsson_variational_2014,invernizzi_rethinking_2020}.

\noindent Atomistic simulations using classical forcefields and electronic structure methods have traditionally relied on software optimized for central processing units (CPUs), as have libraries such as \texttt{PLUMED}.
Over the recent decades, graphics processing unit (GPU) acceleration has become increasingly established in forcefield~\cite{pall_heterogeneous_2020,thompson_lammps_2022,eastman_openmm_2024} and electronic structure simulation codes~\cite{seritan_terachem_2021,li_introducing_2025,iannuzzi_cp2k_2026}.
The availability of GPUs has also supported the widespread adoption of machine learning interatomic potentials (MLIPs), which can approach the accuracy of their first-principles reference calculations at substantially lower computational cost.

\noindent GPU native simulation engines such as \texttt{JAX-MD}~\cite{schoenholz_jax_2020} and \texttt{TorchSim}~\cite{cohen_torchsim_2025} offer further acceleration of forcefield- and MLIP-based simulations through just-in-time (JIT) compilation and batched execution of multiple simulations.
We note that these capabilities are attractive for enhanced sampling methods involving multiple replicas, which in principle can be trivially batched across a fixed number of simulation steps.
Building on the precedent set by codes such as  \texttt{PySAGES}~\cite{zubieta_rico_pysages_2024}, we therefore consider implementations of collective variables (CVs), bias forces, and bias updates that integrate with MLIPs and forcefields but also support batching and end-to-end JIT compilation to improve sampling efficiency and reduce execution time.

\noindent This work makes selected \texttt{PLUMED} functionality available within \texttt{JAX} and documents its validation against \texttt{PLUMED}.
The resulting \texttt{PLUMED-JAX} library includes CVs and two enhanced sampling algorithms, umbrella sampling (US)~\cite{torrie_nonphysical_1977} and (well-tempered) metadynamics (MTD)~\cite{laio_escaping_2002,barducci_well-tempered_2008}, developed using \texttt{gpt-6.1-sol(reasoning\_effort=high)} within the \texttt{Codex} harness.
By using \texttt{PLUMED-JAX} with \texttt{JAX-MD}, we explore through three examples how batching and end-to-end compilation can support enhanced sampling with different physical potentials and computational requirements.
For two lightweight forcefields, we batch simulations on a single GPU and combine this execution with replica exchange between umbrella windows and a shared MTD bias between walkers.
Finally, an application with an MLIP illustrates how \texttt{JAX}'s sharding capabilities can be used to distribute independent umbrella windows across multiple GPUs while retaining batching within each device.

\section{Implementation}
\begin{figure}[tbp]
\centering
\includegraphics[width=\textwidth]{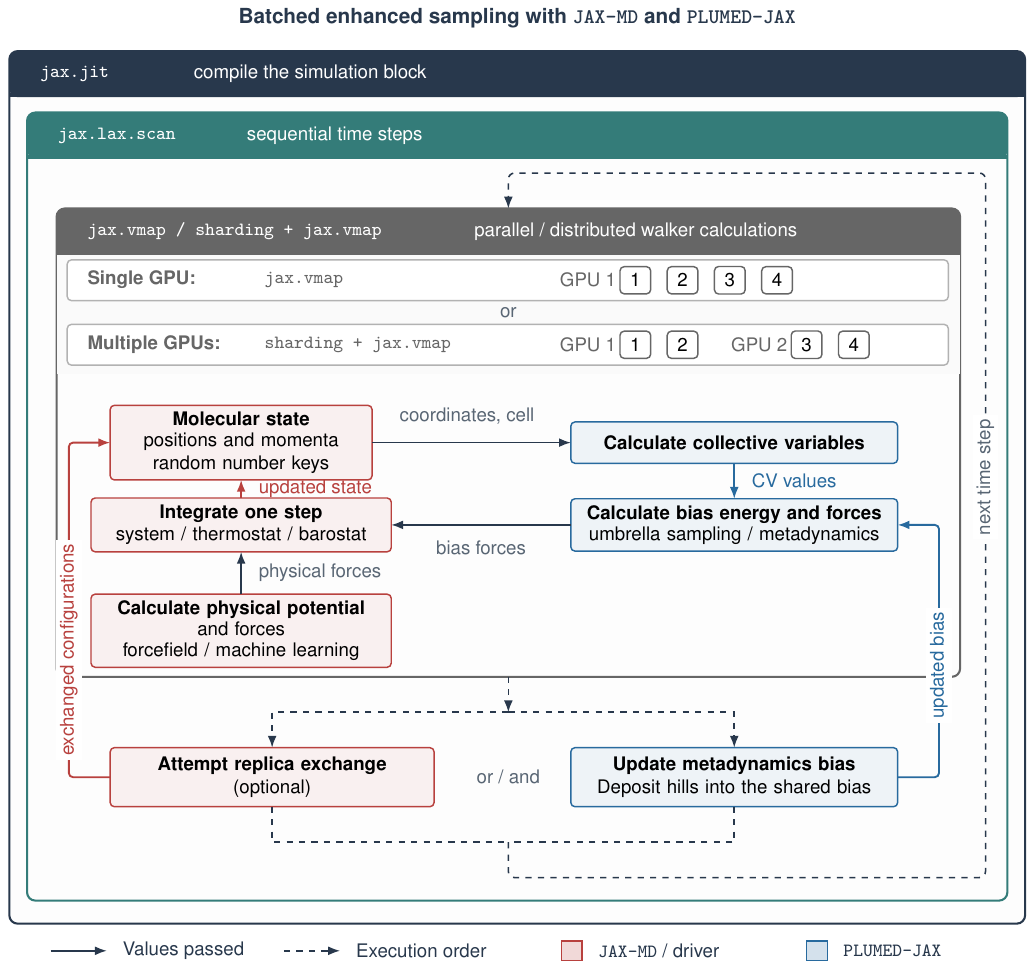}
\caption{Schematic of batched enhanced sampling with \texttt{JAX-MD} and \texttt{PLUMED-JAX}.}
\label{fig:execution}
\end{figure}

\noindent Figure~\ref{fig:execution} summarizes the components of \texttt{PLUMED-JAX}, their interaction with the \texttt{JAX-MD} simulation driver, and the use of \texttt{JAX} utilities for parallel execution and compilation.
We describe the core components, their implementation and design choices, and their validation against \texttt{PLUMED}.

\subsection{Collective variables}

\noindent A collective variable is a function of the atomic coordinates that provides a low-dimensional description of a system's microscopic state.
Mathematically, a CV is simply a scalar function $s\equiv s(\mathbf R,\mathsf h)$, where $\mathbf R$ is the matrix containing the atomic positions as rows and $\mathsf h$ is the matrix containing the three periodic lattice vectors when periodic boundary conditions are implied.

\noindent Each CV is implemented as a \texttt{Python} function using \texttt{JAX} operations, with derivatives with respect to atomic positions and the simulation cell obtained through \texttt{JAX}'s automatic differentiation~\cite{bradbury_jax_2018}.
Appendix~\ref{app:cv-validation} describes the test configurations and validation procedure; Table~\ref{tab:cv-inventory} lists all CV types available in \texttt{PLUMED-JAX}, their scalar components, and the representative option combinations tested.
The same table reports absolute differences from \texttt{PLUMED} in CV values, coordinate derivatives, and derivatives under simultaneous deformation of the coordinates and cell.
The reference calculations use \texttt{PLUMED}~2.10.1 on a CPU, whereas \texttt{PLUMED-JAX} is evaluated on both a CPU and a GPU, using double-precision (64-bit floating-point) arithmetic.

\subsection{Biasing potential}
\noindent Enhanced sampling methods introduce a biasing potential that can depend on time.
The total potential energy felt by the system is
\begin{equation*}
U^{\mathrm{tot}}(\mathbf R,\mathsf h,t)
=U(\mathbf R,\mathsf h)
+U^{\mathrm{bias}}\bigl(\mathbf s(\mathbf R,\mathsf h),t\bigr),
\end{equation*}
where $U(\mathbf R,\mathsf h)$ is the physical potential energy surface, $U^{\mathrm{bias}}(\mathbf s,t)$ is the biasing potential, $\mathbf s(\mathbf R,\mathsf h)=(s_1,\ldots,s_M)$ is the vector of $M$ selected CVs, and $t$ is simulation time.
The total force acting on atom $i$ is $\mathbf F_i^{\mathrm{tot}}=\mathbf F_i+\mathbf F_i^{\mathrm{bias}}$, where $\mathbf F_i=-\nabla_{\mathbf R_i}U$ is the physical force and $\mathbf F_i^{\mathrm{bias}}$ is the additional force from the bias.

\noindent Each bias is implemented as a \texttt{Python} function of the CVs (and time).
\texttt{JAX} obtains the bias forces by differentiating this function composed with the CV functions, using backpropagation to apply the chain rule:
\begin{equation*}
\mathbf F_i^{\mathrm{bias}}
=-\nabla_{\mathbf R_i}U^{\mathrm{bias}}
=-\sum_{\alpha=1}^{M}
\frac{\partial U^{\mathrm{bias}}}{\partial s_\alpha}
\nabla_{\mathbf R_i}s_\alpha.
\end{equation*}
Thus, specifying the bias energy requires no separate implementation of its forces.

\subsubsection{Umbrella sampling}
\noindent US~\cite{torrie_nonphysical_1977} typically uses a harmonic bias to keep the CV vector $\mathbf s$ near a chosen center:
\begin{equation*}
U_a^{\mathrm{bias}}(\mathbf s)
=\frac12(\mathbf s-\mathbf c_a)^{\mathsf T}
\mathbf K_a(\mathbf s-\mathbf c_a),
\end{equation*}
where $a$ labels an umbrella, $\mathbf c_a$ is its center, and $\mathbf K_a$ is a diagonal matrix of force constants for the CVs.
For each periodic CV, the corresponding component of $\mathbf s-\mathbf c_a$ is replaced by the shortest signed difference between the two values.

\noindent Since each bias localizes sampling near its center, US typically uses multiple replicas with centers distributed across the full CV range of interest.
Each replica is called an umbrella.
An advantage of implementing US in \texttt{JAX} is that, when combined with a \texttt{JAX} implementation of the physical potential, as provided by \texttt{JAX-MD} in this work, independent umbrellas can be batched using \texttt{jax.vmap} when they fit on one GPU.
Otherwise, they can be sharded across GPUs, with local batching using \texttt{jax.vmap} on each GPU.

\noindent Simultaneous simulation of the umbrellas also allows replica exchange, which can improve sampling by letting configurations move between neighboring umbrellas.
This combination is known as replica exchange US (REUS)~\cite{sugita_multidimensional_2000}.
For umbrellas $a$ and $b$ at the same temperature, we propose the swap $(\mathbf R_a,\mathbf R_b)\longrightarrow(\mathbf R_b,\mathbf R_a)$, where $\mathbf R_a$ and $\mathbf R_b$ contain the atom positions in the two replicas.
Writing their CV vectors as $\mathbf s_a$ and $\mathbf s_b$, the change in the sum of bias energies and the acceptance probability are
\begin{align*}
\Delta U_{ab}^{\mathrm{bias}}
&=U_a^{\mathrm{bias}}(\mathbf s_b)
 +U_b^{\mathrm{bias}}(\mathbf s_a)
 -U_a^{\mathrm{bias}}(\mathbf s_a)
 -U_b^{\mathrm{bias}}(\mathbf s_b),\\
p_{ab}&=\min\!\left\{1,
\exp\!\left[-\beta\Delta U_{ab}^{\mathrm{bias}}\right]\right\},
\end{align*}
where $\beta$ is the inverse temperature.
The physical energies cancel because all umbrellas use the same physical potential.
On acceptance, positions and momenta are exchanged and the forces are updated for the receiving umbrella.

\noindent These equations describe exchange in the $NVT$ ensemble, however, the  bias and replica exchange is also implemented in the $NPT$ ensemble.
At a common temperature and pressure, the same acceptance rule applies when each cell is exchanged together with its configuration.

\noindent Exchange couples the replicas, so it is performed between blocks of otherwise independent dynamics.
Each block advances the umbrellas for a fixed number of sequential steps using \texttt{jax.lax.scan}, followed by exchange attempts between neighboring umbrellas.
The pairs alternate between $(1,2),(3,4),\ldots$ and $(2,3),(4,5),\ldots$ at successive exchange events, so no umbrella is involved in two simultaneous swaps.
After the exchanges, another block of \texttt{jax.lax.scan} steps begins.

\noindent Appendix~\ref{app:bias-validation} validates US and its REUS variant for a one-dimensional double-well system by comparing the reconstructed free energies with the exact result (Fig.~\ref{fig:double-well-umbrella}).
It also gives the system, simulation setup, and postprocessing methods.

\subsubsection{Metadynamics}
\noindent MTD~\cite{laio_escaping_2002} builds a time-dependent bias by adding localized positive contributions, called Gaussian hills, at the CV values visited during a simulation:
\begin{equation}
U^{\mathrm{bias}}(\mathbf s,t)
=\sum_{n:t_n\leq t}h_n\,g\bigl(q_n(\mathbf s)\bigr),
\qquad
q_n(\mathbf s)
=\frac12(\mathbf s-\mathbf s_n)^{\mathsf T}
\boldsymbol\Sigma^{-1}(\mathbf s-\mathbf s_n),
\label{eq:MTD-bias}
\end{equation}
where $n$ labels a hill, $t_n$ is the time at which it is added, $h_n$ is its height, and $g$ is an unnormalized Gaussian, shifted and rescaled to vanish at a finite cutoff.
Its argument $q_n(\mathbf s)$ is half the squared distance between the current CV vector $\mathbf s$ and the hill center $\mathbf s_n$, measured in units of the respective hill widths.
The diagonal matrix $\boldsymbol\Sigma$ contains the squared hill widths, with entry $\Sigma_{\alpha\alpha}=\sigma_\alpha^2$ for CV component $\alpha$.
For periodic CVs, the corresponding components of $\mathbf s-\mathbf s_n$ are replaced by the shortest signed differences.
When the dynamics of the system is biased by this potential, the hills raise the energy of previously visited regions over time, encouraging exploration of new regions in CV space.

\noindent Ordinary MTD adds hills of a constant height $h_0$.
Well-tempered MTD~\cite{barducci_well-tempered_2008} instead reduces the height in regions where the bias has already accumulated:
\begin{equation*}
h_n=h_0\exp\!\left[-\frac{\beta U^{\mathrm{bias}}(\mathbf s_n,t_n^-)}
{\gamma-1}\right],
\end{equation*}
where $h_0$ is the initial hill height, $t_n^-$ denotes the instant before the new hill is added, and $\gamma>1$ is the bias factor that controls how strongly the height decreases.

\noindent The original MTD algorithm builds its bias along one trajectory and is therefore serial in time, but it can be parallelized using multiple walker MTD~\cite{raiteri_efficient_2006}.
In this approach, several replicas, called walkers, explore the system simultaneously while contributing hills to one shared bias.
The sum in Eq.~\eqref{eq:MTD-bias} then includes hills from all walkers, so a region visited by one walker contributes to the bias felt by the others.
With a \texttt{JAX} physical potential, the walkers can use the same single-GPU batching or multi-GPU sharding with local \texttt{jax.vmap} described for umbrellas, with a shared bias update.

\noindent Between hill additions, the walkers advance independently for a fixed number of sequential steps using \texttt{jax.lax.scan}.
At the end of each block, every walker contributes a hill centered at its current CV vector.
All (well-tempered) heights are calculated from the same bias before these hills are added, after which the shared bias and the forces are updated and the next \texttt{jax.lax.scan} block begins.
Hill centers and heights are stored in arrays whose sizes are chosen before compilation.

\noindent Evaluating the sum directly becomes more expensive as the number of hills grows.
An optional grid implementation instead stores the bias on a regular CV grid and evaluates it by interpolation between nearby grid points, following established grid representations of the bias~\cite{bonomi_plumed_2009}.
At fixed grid resolution, this evaluation cost is independent of the number of deposited hills.

\noindent Appendix~\ref{app:bias-validation} validates single walker and multiple walker well-tempered MTD against the exact free energy of the same double-well system (Fig.~\ref{fig:double-well-MTD}).
It also gives the simulation setup and postprocessing methods.

\section{Results}
\noindent We now turn to three realistic systems studied in previous work, providing published free energy profiles for comparison.
These examples allow us to test the implementation beyond the controlled benchmarks and demonstrate how \texttt{JAX} batching and execution across multiple GPUs can be used with different physical potentials and sampling methods.

\subsection{Replica exchange umbrella sampling: butane torsional free energy surface}

\begin{figure}[tbp]
\centering
\includegraphics[width=0.65\linewidth]{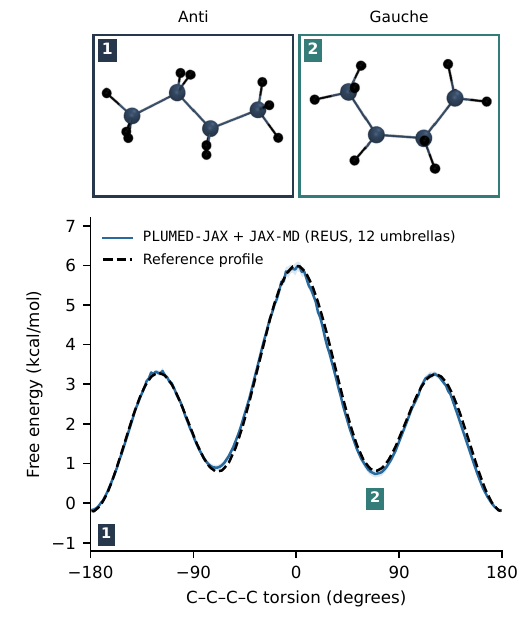}
\caption{Anti (1) and gauche (2) butane conformers and the torsional free energy at 300~K using the optimized potentials for liquid simulations forcefield.
Replica exchange umbrella samples are combined using the weighted histogram analysis method.
The reference profile is digitized from the spectral adaptive biasing force calculation in Reference~\citenum{zubieta_rico_pysages_2024}.
Shading is one standard error. 
Snapshots are rendered with \texttt{VMD}~\cite{humphrey_vmd_1996}.
}
\label{fig:butane}
\end{figure}

\noindent We first consider the torsional free energy profile of butane.
Rotation about its central carbon--carbon bond connects the anti (trans) conformer, near a C--C--C--C torsion of $\pm180^\circ$, with two equivalent gauche conformers near $\pm60^\circ$~\cite{jorgensen_development_1996}.
The intervening barriers make these regions less frequently visited than the minima in unbiased dynamics, providing a simple test of sampling across the full torsional range.
In Ref.~\citenum{zubieta_rico_pysages_2024}, this profile was calculated using a spectral adaptive biasing force method~\cite{zubieta_rico_sobolev_2022} implemented in \texttt{PySAGES}  with the optimized potentials for liquid simulations (\texttt{OPLS}) forcefield.
Here we use REUS and take the curve digitized from Fig.~12 of Ref.~\citenum{zubieta_rico_pysages_2024} as the reference profile.

\noindent We use the \texttt{JAX-MD} implementation of the \texttt{OPLS} forcefield with parameters from the released input of Ref.~\citenum{zubieta_rico_pysages_2024} for flexible, 14-atom butane in vacuum, with the same temperature of 300~K and cubic cell of side $41\,\angstrom$.
The nonbonded cutoff is $12\,\angstrom$, Lennard--Jones parameters use geometric mixing, and interactions between atoms separated by three bonds are scaled by 1 for Lennard--Jones and $1/2$ for Coulomb energies.
Our electrostatic settings differ from the released reference.
We use Ewald electrostatics with splitting parameter $0.23\,\angstrom^{-1}$ and reciprocal lattice indices from $-10$ to $10$ in each direction, whereas the reference uses a mesh solver.

\noindent Our REUS calculation uses twelve periodic umbrella biases on the C--C--C--C torsion, each with force constant $5\,\kcalmol\,\mathrm{rad}^{-2}$, and batches all umbrellas on one A100 GPU.
Each umbrella undergoes 50~ps of equilibration and 1.00~ns of production, with exchanges attempted between neighboring umbrellas every 20 steps.
Of all proposed exchanges, 41.7\% are accepted.

\noindent We reconstruct the free energy profile with the weighted histogram analysis method~\cite{kumar_weighted_1992}, without averaging its two symmetric halves (Fig.~\ref{fig:butane}).
The barrier at a torsion of $0^\circ$ relative to the anti minimum is $6.16\pm0.122\,\kcalmol$, compared with $6.19\,\kcalmol$ in the reference, while the gauche$^{-}$ minimum lies $1.08\pm0.0714\,\kcalmol$ above anti, compared with $1.01\,\kcalmol$.
After an additive energy shift, the root-mean-square (RMS) difference between the profiles is $0.104\,\kcalmol$; profiles from the first and second trajectory halves differ by $0.107\,\kcalmol$ RMS.
Errors are estimated from ten 100~ps blocks, leaving out one block at a time (block jackknife).
These results demonstrate excellent agreement between REUS as implemented in \texttt{PLUMED-JAX} and the reference profile.

\subsection{Multiple walker metadynamics: silicon crystallization}

\begin{figure}[tbp]
\centering
\includegraphics[width=0.65\linewidth]{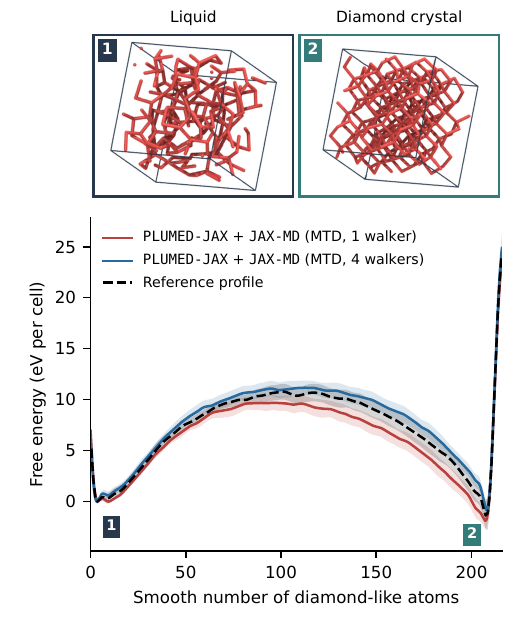}
\caption{Liquid (1) and diamond crystal (2) silicon configurations, and free energy profiles for the \texttt{Stillinger-Weber} potential at 1700~K and a target pressure of 1~bar.
Profiles are estimated by scaling the final well-tempered metadynamics bias, using one or four walkers.
The reference profile is the released \texttt{LAMMPS}/\texttt{PLUMED} result~\cite{piaggi_silicon_tutorial}.
Shading shows the standard deviation of eleven late-time bias profiles.
Snapshots are rendered with \texttt{VMD}~\cite{humphrey_vmd_1996}.
}
\label{fig:silicon}
\end{figure}

\noindent We next consider crystallization of silicon, in which atoms rearrange from a disordered liquid into an ordered diamond crystal.
In contrast to the isolated-molecule butane example, this is a bulk periodic system, and its CV depends on all atoms rather than on a four-atom torsion.
Ref.~\citenum{piaggi_silicon_tutorial} provides a single walker well-tempered MTD calculation using \texttt{LAMMPS} and \texttt{PLUMED} with the \texttt{Stillinger-Weber} potential~\cite{stillinger_silicon_1985}.
We use its released numerical free energy profile as the reference and compare single walker and multiple walker calculations at equal total sampling time.

\noindent We simulate 216 silicon atoms using the \texttt{JAX-MD} implementation of the \texttt{Stillinger-Weber} potential at 1700~K and a target pressure of 1~bar, starting from a diamond crystal in a cubic cell of side $16.2921\,\angstrom$.
Following the reference, the CV is a smooth count of atoms whose neighbor arrangements resemble those in a diamond crystal, so it is small in the liquid and approaches 216 in the crystal.
Each atom's neighbors are compared with two tetrahedral reference arrangements representing the two environments in the diamond lattice~\cite{piaggi_phase_2019}.
The reference lattice constant is $5.431\,\angstrom$, and a Gaussian width of $0.400\,\angstrom$ controls how much the neighbor positions can deviate from these arrangements.
We combine the two similarity scores using a smooth maximum with coefficient 100, then turn the score into a continuous count using a threshold of 0.500 and a switching function with powers 6 and 12.

\noindent The timestep is 2~fs, and preparation comprises 50~ps at constant volume followed by 50~ps at constant pressure, as in the reference.
Constant pressure dynamics use an isotropic barostat with a timescale of 1~ps, which expands or contracts the cell by the same factor in each direction.
We thermalize the particle and volume momenta using Langevin updates with friction $1\,\mathrm{ps}^{-1}$; the reference instead uses stochastic velocity rescaling~\cite{bussi_isothermal_2009} with a 0.1~ps thermostat timescale and a 1~ps pressure timescale.

\noindent We retain the reference well-tempered MTD settings with an initial hill height of $0.622$~eV, a width of 3.00 in the count of crystalline atoms, bias factor 50, and one hill per walker every 1~ps.
We compare one walker run for 5.00~ns with four walkers run for 1.25~ns each, batched on an A100 GPU and contributing to a shared bias.
Both calculations accumulate 5.00~ns of sampling and 5,000 hills.

\noindent Both calculations repeatedly visit the liquid and crystal states (Fig.~\ref{fig:silicon}).
We estimate the free energy from the final well-tempered bias, multiplying its negative by $\gamma/(\gamma-1)$, where $\gamma=50$.
After alignment at the liquid minimum, the RMS differences from the reference are 1.09~eV per cell for one walker and 0.656~eV per cell for four walkers.
For four walkers, the liquid-to-solid barrier is $11.2\pm0.669$~eV per cell, compared with $10.8\pm0.721$~eV per cell in the reference, and the solid minimum lies $0.829\pm1.02$~eV per cell below the liquid minimum, compared with $1.36\pm1.36$~eV per cell.
The shading and quoted spreads are standard deviations of eleven free energy profiles calculated at late times in each run.
Overall, both calculations recover the liquid and crystal minima and a barrier that agrees with the reference within the observed late time spread.

\begin{figure}[tbp]
\centering
\includegraphics[width=0.65\linewidth]{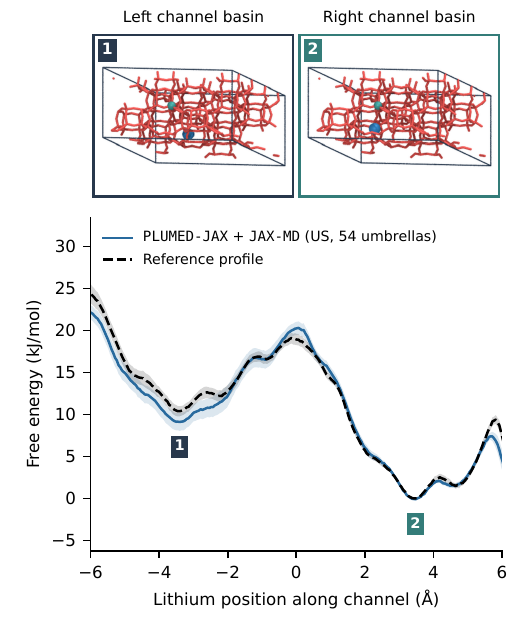}
\caption{Starting structures released by the authors near the two channel minima (1,2), with lithium in blue, and the lithium free energy profile.
Umbrella sampling uses the converted \texttt{MACE-MP-0-medium} potential, and profiles are reconstructed using the multistate Bennett acceptance ratio.
The reference profile is reconstructed from the authors' released umbrella data~\cite{majumdar_lithium_data_2026}.
Water is omitted from the images; all 56 molecules are present in the simulations.
Shading is one standard error.
Snapshots are rendered with \texttt{VMD}~\cite{humphrey_vmd_1996}.
}
\label{fig:lithium}
\end{figure}

\subsection{Distributed umbrella sampling with a machine learning potential: lithium transport in hydrated chabazite}

\noindent Finally, we consider lithium transport through hydrated chabazite, a porous crystal in which ions move between favorable sites inside water-filled channels.
The free energy barriers for this motion depend on how the ion interacts with water and the surrounding crystal, as studied in Ref.~\citenum{majumdar_transport_2026} using the \texttt{MACE-MP-0-medium} machine learning potential and US.
This example combines a larger solvated system with a machine learning potential, allowing us to demonstrate how umbrellas can be distributed across multiple GPUs while retaining batching within each device.
We use the authors' released umbrella data to reconstruct the reference profile~\cite{majumdar_lithium_data_2026}.

\noindent We use the released 601-atom structure containing one lithium ion and 56 water molecules, retaining its fixed triclinic cell.
The CV measures the lithium position along a channel.
We define the channel axis using the mean positions of two eight-atom groups in the crystal framework, and project the lithium position onto this axis relative to the midpoint between the groups.
Negative and positive CV values therefore describe positions on opposite sides of that midpoint.

\noindent The physical energy and forces are evaluated with the converted \texttt{MACE-MP-0-medium} model~\cite{batatia_mace_2022,batatia_foundation_2025} using the \texttt{MACE-JAX} implementation~\cite{mace_jax_software}, while \texttt{PLUMED-JAX} supplies the umbrella bias.
The model has two interaction layers, 128 scalar and 128 vector features, and a $6\,\angstrom$ cutoff; it uses \texttt{cuEquivariance} (\texttt{cuEq})~\cite{nvidia_cuequivariance} for GPU acceleration.
Model evaluations use 32-bit floating-point arithmetic, while integration and energy accumulation use 64-bit arithmetic.

\noindent We run 54 independent harmonic umbrellas spaced by $0.250\,\angstrom$, with force constant $89.8\,\kjmol\,\angstrom^{-2}$ and a target temperature of 450~K.
Each umbrella runs for 0.800~ns with a 1~fs timestep and a Langevin thermostat with friction $1\,\mathrm{ps}^{-1}$; the first 2~ps are discarded.
The umbrella settings and duration match the released reference data.
\texttt{JAX} sharding distributes complete umbrellas across three A100 GPUs, with \texttt{jax.vmap} batching 18 umbrellas on each GPU.
The umbrellas evolve independently, without replica exchange.

\noindent We reconstruct both our profile and the reference using the multistate Bennett acceptance ratio~\cite{shirts_statistically_2008}.
After an additive energy shift, the RMS difference from the reference is $0.984\,\kjmol$ (Fig.~\ref{fig:lithium}).
The barrier from the left minimum toward the right is $11.2\pm0.757\,\kjmol$, compared with $8.71\pm0.491\,\kjmol$ in the reference, giving a difference of $2.48\pm0.902\,\kjmol$.
The right minimum lies $9.21\pm1.05\,\kjmol$ below the left, compared with $10.4\pm0.616\,\kjmol$ in the reference.
Errors are estimated by block averaging with eight approximately 100~ps blocks, keeping the positions of the minima and barrier fixed at those found in each full profile.
Noting that these errors are biased (and thus underestimated), it may be concluded that the profiles show fair agreement in the relative stability of the two sites and the barrier between them.

\subsection{Computational performance}
\begin{figure}[tbp]
\centering
\includegraphics[width=\linewidth]{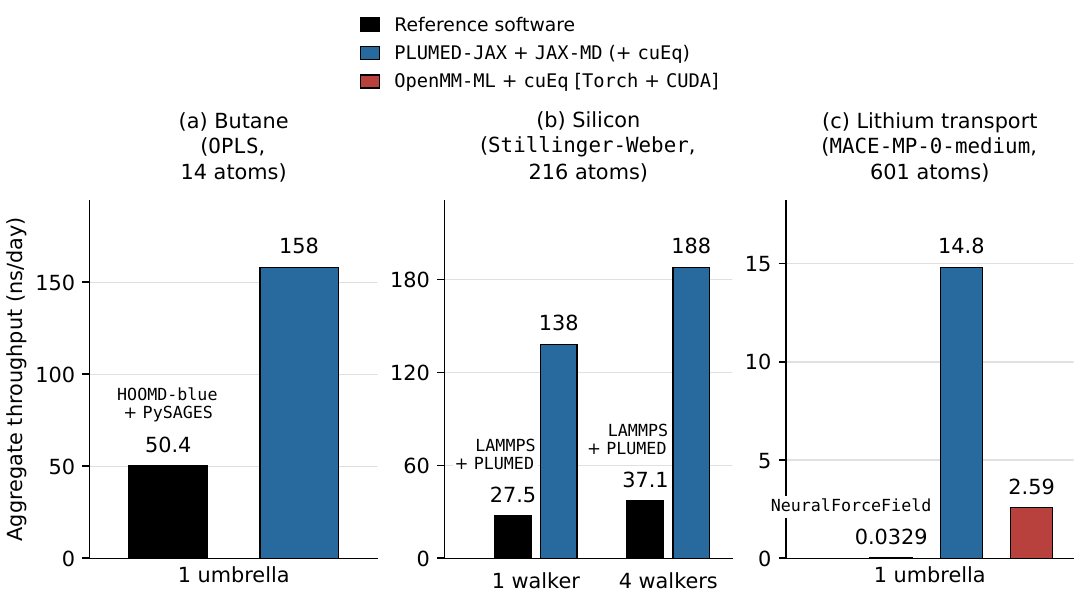}
\caption{Molecular dynamics throughput on one RTX A6000, using separate linear y-axes for the three systems.
Butane and lithium use one umbrella; silicon includes one and four well-tempered metadynamics walkers, with rates summed across walkers.
The potentials and atom counts are shown above each panel; the reference setups follow the butane~\cite{zubieta_rico_pysages_2024}, silicon~\cite{piaggi_silicon_tutorial}, and lithium~\cite{majumdar_lithium_data_2026} protocols described in the text.
The \texttt{JAX} and \texttt{OpenMM-ML} timings for lithium both use \texttt{cuEq}.
}
\label{fig:a6000-throughput}
\end{figure}

\noindent We next assess the computational cost of these calculations by measuring how much molecular dynamics (MD) can be performed in one day of wall time on one NVIDIA RTX A6000 (Fig.~\ref{fig:a6000-throughput}).
Each timing covers complete MD steps which includes building or updating neighbor lists, evaluating the physical forces and the CVs, calculating the bias forces and updating the bias when required, and advancing the positions, momenta, and temperature or pressure control.
Setup, compilation, and warm-up are excluded.
Butane and lithium use one umbrella, while silicon uses either one or four walkers.
For several walkers, we report aggregate throughput, meaning the sum of their simulated times divided by wall time.

\noindent For butane, the reference combines \texttt{HOOMD-blue}~\cite{anderson_hoomd_2020}, which evaluates the forcefield and advances the dynamics, with \texttt{PySAGES}, which supplies the harmonic umbrella bias using the released butane setup of Ref.~\citenum{zubieta_rico_pysages_2024}.
For silicon, \texttt{LAMMPS} evaluates the \texttt{Stillinger-Weber} forces using its GPU implementation (\texttt{Kokkos}/\texttt{CUDA}) and advances the dynamics, while \texttt{PLUMED} evaluates the CV and MTD bias.
For lithium, we time the authors' released \texttt{NeuralForceField} (\texttt{NFF}) driver with \texttt{MACE-MP-0-medium}~\cite{majumdar_lithium_data_2026}, and also \texttt{OpenMM-ML}, as community code that supports enhanced sampling with MLIPs, with \texttt{MACE-MP-0-medium} evaluated through \texttt{PythonForce} \texttt{PyTorch} \texttt{MACE} adapter~\cite{paszke_pytorch_2019} and \texttt{cuEq}.
In the latter calculation, \texttt{OpenMM} evaluates the umbrella bias and advances the dynamics using \texttt{CUDA} kernels.
The lithium \texttt{JAX} and \texttt{OpenMM-ML} calculations both use 32-bit arithmetic for \texttt{MACE-MP-0-medium}, 64-bit arithmetic for the bias and MD integration, and \texttt{cuEq} acceleration.
The reference configurations are the fastest successful settings tested, using the existing software implementations.

\noindent The timesteps are 0.5~fs for butane, 2~fs for silicon, and 1~fs for lithium.
\texttt{HOOMD} uses a $1\,\angstrom$ neighbor list buffer, and \texttt{LAMMPS} uses eight CPU threads for one silicon walker or four threads per walker for four walkers.
Both silicon implementations use a bias grid initialized with 4,500 hills.

\noindent The complete \texttt{JAX} simulations are 3.14 times faster for butane and 5.02 times faster for single walker silicon than their measured reference configurations.
Batching four silicon walkers sharing a bias increases aggregate \texttt{JAX} throughput from 138 to 188~ns/day, a factor of 1.36, and gives 5.06 times the throughput of the four-walker \texttt{LAMMPS}/\texttt{PLUMED} calculation.
For lithium, the compiled \texttt{JAX} simulation reaches 14.8~ns/day, compared with 2.59~ns/day for \texttt{OpenMM-ML}.
The \texttt{JAX} rate is the one-window batching control in Appendix~\ref{app:window-scaling}.
This comparison illustrates the importance of how the force model is connected to the simulation driver: the measured \texttt{OpenMM} \texttt{MACE} adapter builds the graph on the CPU and transfers inputs and forces between CPU and GPU at every step, whereas the compiled \texttt{JAX} loop keeps graph construction, force evaluation, biasing, and integration on the GPU.
We also attempted a like-for-like comparison using the same \texttt{MACE-JAX} model within \texttt{OpenMM-JAX} (\texttt{openmmjax-cu13} version 0.1.4).
The export interface does not support the \texttt{cuEq} operation used by this model, preventing this comparison.

\noindent To understand how several lithium umbrellas share the GPU, we compare batching with \texttt{NVIDIA Multi-Process Service} (\texttt{MPS}), which allows GPU work from separate processes to overlap (Appendix~\ref{app:window-scaling}).
For \texttt{OpenMM}, this allows CPU graph construction and data transfers to overlap with GPU work from other umbrellas; for \texttt{JAX}, separate processes under \texttt{MPS} do not improve on the measured batched throughput.
This behavior is consistent with effective GPU use by the compiled \texttt{JAX} simulation: sharing the device among more umbrellas reduces the rate of each trajectory while providing a modest increase in their combined rate.
Together, these comparisons demonstrate the benefit of compiling the complete simulation loop and keeping its data on the GPU.

\section{Conclusions}
\noindent \texttt{PLUMED-JAX} makes a small subset of \texttt{PLUMED}'s capabilities available to \texttt{JAX}-based simulation workflows.
Comparisons with \texttt{PLUMED} establish numerical agreement for the tested CV values, derivatives, and bias calculations, while a one-dimensional double well provides an exact free energy reference for validating the sampling methods.
Combining these components with \texttt{JAX-MD} allows force evaluation, integration, and bias updates to remain within a compiled calculation.
The applications to butane and silicon demonstrate batched US with replica exchange and multiple walker MTD with a shared bias on one GPU, while lithium transport illustrates how US with a machine learning potential can be distributed across multiple GPUs.
The A6000 benchmarks show the computational benefit of this approach: batching four silicon walkers increases aggregate throughput by a factor of 1.36, while lithium dynamics with one umbrella reaches 14.8~ns/day, 5.71 times the throughput of the tested \texttt{OpenMM-ML} configuration using the same \texttt{MACE-MP-0-medium} checkpoint, model precision, and \texttt{cuEq} acceleration.
For lithium, a diagnostic comparison using \texttt{JAX}+\texttt{MPS} did not improve on batching, consistent with effective GPU use by the compiled simulation.
Together, numerical validation and these execution comparisons support the use of \texttt{PLUMED-JAX} with \texttt{JAX-MD} to combine enhanced sampling, dynamics performed entirely on the GPU, and parallel simulations within a common framework.

\section*{Acknowledgments}
\noindent The author thanks \texttt{PLUMED}'s developers and contributors for developing and maintaining the collective variables and enhanced sampling methods adapted here, and for making their implementations and documentation openly available for experimentation.
\noindent The author thanks all members of the Initiative for Computational Catalysis at the Flatiron Institute for insightful discussions and Abhijeet Gangan for stimulating discussions.
Computational resources were provided by the Flatiron Institute, a division of the Simons Foundation.

\section*{Data and software availability}
\noindent The data supporting this work will be made available at a later stage.
\texttt{PLUMED-JAX} is available under the Apache~2.0 license at \url{https://github.com/venkatkapil24/plumed-jax}.
The \texttt{JAX-MD} branch used in this work is available at \url{https://github.com/venkatkapil24/jax-md/tree/working-branch}.

\clearpage
\appendix
\setcounter{figure}{0}
\renewcommand{\thefigure}{A\arabic{figure}}
\renewcommand{\theHfigure}{appendix.\arabic{figure}}
\setcounter{table}{0}
\renewcommand{\thetable}{A\arabic{table}}
\renewcommand{\theHtable}{appendix.\arabic{table}}
\section{Numerical validation}
\label{app:cv-validation}

\subsection{Validation of Collective Variables}
\noindent To test the collective variable (CV) implementations, we generate nine artificial point clouds containing 5, 16, or 64 atoms, using three fixed random seeds (1701, 2903, and 4309) per size.
Positions are drawn uniformly from a $4\,\angstrom$ cube, with pair separations above $0.45\,\angstrom$ and noncollinear atom selections for torsions.

\noindent Each cloud is tested in four cases: without periodic boundaries, in a cubic cell of side $12\,\angstrom$, in an orthorhombic cell of side lengths $(12,13,14)\,\angstrom$, and in a triclinic cell with row vectors $(12,0,0)$, $(4.3,11.7,0)$, and $(-1.1,2.4,13.2)\,\angstrom$.
This gives 36 configurations.
For periodic tests, the cloud is shifted by $(10.1,11.2,12.3)\,\angstrom$ and brought back inside the cell using the periodic boundaries.

\noindent We construct 19 further five-atom geometries by placing selected atoms at specified distances and angles to test cases unlikely to occur in random clouds.
These test contact functions near a finite $0/0$ limit and where they become zero (six geometries), torsions near the change from $+\pi$ to $-\pi$ (two), nearly straight bond angles (two), and distances where a restraining potential begins to act (three).
Two geometries test distances across periodic boundaries, using a $5\,\angstrom$ cube or a tilted cell with row vectors $(5,0,0)$, $(4.6,1.4,0)$, and $(0.5,0.7,4.8)\,\angstrom$.
Four geometries test reference-based CVs: an exact reference, a distorted and rotated reference, and the two endpoints of a reference path.

\noindent For each applicable CV $s(\mathbf R,\mathsf h)$, we compare its value and the following derivatives, called \emph{gradient} and \emph{deformation} in Table~\ref{tab:cv-inventory}:
\begin{equation*}
\underbrace{\left.\frac{\partial s}{\partial R_{i\alpha}}
\right|_{\mathsf h}}_{\text{gradient}},
\qquad
\underbrace{\left.
\frac{\partial s\bigl(\mathbf R(\mathbf I+\boldsymbol\epsilon),
\mathsf h(\mathbf I+\boldsymbol\epsilon)\bigr)}
{\partial\epsilon_{\alpha\beta}}
\right|_{\boldsymbol\epsilon=\mathbf0}}_{\text{deformation}}.
\end{equation*}
The gradient measures the change in the CV when an atom's position changes with the cell fixed; $R_{i\alpha}$ is component $\alpha$ of the position of atom $i$.
The deformation derivative measures its change when the positions and cell are scaled or sheared together: $\mathbf R$ contains atom positions as rows, $\mathbf I$ is the $3\times3$ identity matrix, $\boldsymbol\epsilon$ is a dimensionless $3\times3$ deformation matrix, and $\alpha,\beta$ run over $x,y,z$.
For nonperiodic tests, the cell argument is omitted.

\noindent The 55 configurations give 3,422 comparisons per device, covering 95 scalar outputs and option combinations; each quantity is tested only on applicable configurations.
\texttt{PLUMED-JAX} runs on a central processing unit (CPU) and an NVIDIA RTX A6000 graphics processing unit (GPU), using \texttt{JAX}~0.10.2; the reference is \texttt{PLUMED}~2.10.1 on a CPU, with all comparisons using 64-bit floating-point arithmetic.

\noindent Table~\ref{tab:cv-inventory} gives the largest absolute differences from \texttt{PLUMED} over the tested configurations and, for derivatives, over their individual components.
The row unit applies to value and deformation errors; gradient errors have that unit per angstrom.
Summed coordination errors are reported per input atom; other quantities are not divided by atom count.
The symbol $1$ means dimensionless; ``numeric'' marks expressions without a single physical unit, evaluated with distances in angstroms.
RMSD denotes root-mean-square deviation.
Path coordinates and bond orientational order use the definitions of References~\citenum{branduardi_path_2007} and \citenum{steinhardt_order_1983}, respectively.
\label{app:cv-text-end}

\clearpage

\begin{table}[tbp]
\centering
\small
\setlength{\tabcolsep}{3pt}
\renewcommand{\arraystretch}{1.10}
\caption{Maximum absolute errors for 95 scalar collective variable components, option combinations, switching functions, composition operations, and bias energies against \texttt{PLUMED}~2.10.1, using 64-bit floating-point arithmetic on a central processing unit and an NVIDIA RTX A6000 graphics processing unit.
Horizontal rules separate collective variable types or operation families.}
\label{tab:cv-inventory}
\resizebox{\linewidth}{!}{%
\begin{tabular}{@{}>{\raggedright\arraybackslash}p{0.32\textwidth}rrrrrr@{}}
\noalign{\hrule height .6pt\vskip 3pt}
\textbf{Scalar quantity / option (unit)} & \multicolumn{3}{c}{\textbf{CPU}} & \multicolumn{3}{c}{\textbf{GPU}} \\

 & Value & Gradient & Deformation & Value & Gradient & Deformation \\
\noalign{\vskip 2pt\hrule height .35pt\vskip 2pt}
Distance [$\angstrom$] & \num{1.03e-13} & \num{5.17e-14} & \num{1.19e-13} & \num{1.03e-13} & \num{5.17e-14} & \num{1.19e-13} \\
\noalign{\vskip 2pt\hrule height .35pt\vskip 2pt}
Position $x$ [$\angstrom$] & \num{9.24e-14} & \(0\) & \num{9.24e-14} & \num{9.24e-14} & \(0\) & \num{9.24e-14} \\
Position $y$ [$\angstrom$] & \num{9.24e-14} & \(0\) & \num{9.24e-14} & \num{9.24e-14} & \(0\) & \num{9.24e-14} \\
Position $z$ [$\angstrom$] & \num{8.70e-14} & \(0\) & \num{9.24e-14} & \num{8.70e-14} & \(0\) & \num{9.24e-14} \\
\noalign{\vskip 2pt\hrule height .35pt\vskip 2pt}
Center of mass--atom distance [$\angstrom$] & \num{1.06e-13} & \num{1.19e-13} & \num{1.15e-13} & \num{1.05e-13} & \num{1.19e-13} & \num{1.15e-13} \\
Center of mass $x$ [$\angstrom$] & \num{4.73e-14} & \num{1.44e-16} & \num{5.86e-14} & \num{4.73e-14} & \num{1.44e-16} & \num{5.86e-14} \\
Center of mass $y$ [$\angstrom$] & \num{5.95e-14} & \num{1.44e-16} & \num{5.86e-14} & \num{5.95e-14} & \num{1.44e-16} & \num{5.86e-14} \\
Center of mass $z$ [$\angstrom$] & \num{4.62e-14} & \num{1.44e-16} & \num{5.86e-14} & \num{4.62e-14} & \num{1.44e-16} & \num{5.86e-14} \\
\noalign{\vskip 2pt\hrule height .35pt\vskip 2pt}
Bond angle [$\mathrm{rad}$] & \num{6.46e-14} & \num{3.80e-11} & \num{7.44e-14} & \num{6.46e-14} & \num{3.80e-11} & \num{7.44e-14} \\
\noalign{\vskip 2pt\hrule height .35pt\vskip 2pt}
Signed torsion [$\mathrm{rad}$] & \num{1.15e-13} & \num{1.08e-13} & \num{9.65e-14} & \num{1.15e-13} & \num{1.08e-13} & \num{9.65e-14} \\
\noalign{\vskip 2pt\hrule height .35pt\vskip 2pt}
Proton transfer coordinate [$\angstrom$] & \num{1.73e-13} & \num{5.97e-14} & \num{1.81e-13} & \num{1.73e-13} & \num{5.95e-14} & \num{1.81e-13} \\
\noalign{\vskip 2pt\hrule height .35pt\vskip 2pt}
Gyration radius, equal weights [$\angstrom$] & \num{3.11e-14} & \num{1.02e-14} & \num{6.49e-14} & \num{3.13e-14} & \num{1.02e-14} & \num{6.42e-14} \\
Gyration radius, mass weights [$\angstrom$] & \num{4.20e-14} & \num{1.66e-14} & \num{5.97e-14} & \num{4.22e-14} & \num{1.65e-14} & \num{6.02e-14} \\
\noalign{\vskip 2pt\hrule height .35pt\vskip 2pt}
Coordination, two groups [$\mathrm{atom}^{-1}$] & \num{2.67e-14} & \num{3.28e-14} & \num{5.55e-14} & \num{2.66e-14} & \num{3.28e-14} & \num{5.55e-14} \\
Coordination, one group [$\mathrm{atom}^{-1}$] & \num{4.17e-14} & \num{3.50e-14} & \num{5.74e-14} & \num{4.09e-14} & \num{3.49e-14} & \num{5.74e-14} \\
Coordination, overlap [$\mathrm{atom}^{-1}$] & \num{2.77e-14} & \num{3.28e-14} & \num{5.53e-14} & \num{2.77e-14} & \num{3.28e-14} & \num{5.53e-14} \\
Coordination, shifted [$\mathrm{atom}^{-1}$] & \num{2.64e-14} & \num{2.13e-14} & \num{3.22e-14} & \num{2.63e-14} & \num{2.13e-14} & \num{3.22e-14} \\
Coordination, general [$\mathrm{atom}^{-1}$] & \num{2.40e-14} & \num{4.07e-12} & \num{5.90e-12} & \num{2.40e-14} & \num{4.07e-12} & \num{5.90e-12} \\
\noalign{\vskip 2pt\hrule height .35pt\vskip 2pt}
Pair contact (1,2), raw [$1$] & \num{5.37e-14} & \num{6.62e-14} & \num{1.28e-13} & \num{5.37e-14} & \num{6.62e-14} & \num{1.28e-13} \\
Pair contact (1,2), stretched [$1$] & \num{5.37e-14} & \num{6.62e-14} & \num{1.28e-13} & \num{5.37e-14} & \num{6.62e-14} & \num{1.28e-13} \\
Pair contact (2,3), raw [$1$] & \num{7.27e-14} & \num{1.33e-13} & \num{2.17e-13} & \num{7.27e-14} & \num{1.33e-13} & \num{2.17e-13} \\
Pair contact (2,3), stretched [$1$] & \num{7.27e-14} & \num{1.33e-13} & \num{2.17e-13} & \num{7.27e-14} & \num{1.33e-13} & \num{2.17e-13} \\
Pair contact (1,4), raw [$1$] & \num{2.26e-14} & \num{5.40e-14} & \num{6.45e-14} & \num{2.26e-14} & \num{5.40e-14} & \num{6.45e-14} \\
Pair contact (1,4), stretched [$1$] & \num{2.26e-14} & \num{5.40e-14} & \num{6.45e-14} & \num{2.26e-14} & \num{5.40e-14} & \num{6.45e-14} \\
\noalign{\vskip 2pt\hrule height .35pt\vskip 2pt}
Matrix, one group, $M_{11}$ [$1$] & \(0\) & \(0\) & \(0\) & \(0\) & \(0\) & \(0\) \\
Matrix, one group, $M_{12}$ [$1$] & \num{5.37e-14} & \num{6.62e-14} & \num{1.28e-13} & \num{5.37e-14} & \num{6.62e-14} & \num{1.28e-13} \\
Matrix, one group, $M_{21}$ [$1$] & \num{5.37e-14} & \num{6.62e-14} & \num{1.28e-13} & \num{5.37e-14} & \num{6.62e-14} & \num{1.28e-13} \\
Matrix, one group, $M_{22}$ [$1$] & \(0\) & \(0\) & \(0\) & \(0\) & \(0\) & \(0\) \\
Matrix, one group, \texttt{EXP} [$1$] & \num{2.03e-14} & \num{1.87e-14} & \num{2.55e-14} & \num{2.03e-14} & \num{1.87e-14} & \num{2.55e-14} \\
Matrix, one group, \texttt{GAUSSIAN} [$1$] & \num{4.24e-14} & \num{2.21e-14} & \num{5.01e-14} & \num{4.24e-14} & \num{2.21e-14} & \num{5.01e-14} \\
Matrix, one group, \texttt{TANH} [$1$] & \num{2.28e-14} & \num{2.27e-14} & \num{2.90e-14} & \num{2.27e-14} & \num{2.27e-14} & \num{2.90e-14} \\
Matrix, one group, \texttt{SMAP} [$1$] & \num{4.20e-14} & \num{5.90e-14} & \num{1.08e-13} & \num{4.20e-14} & \num{5.90e-14} & \num{1.08e-13} \\
Matrix, one group, \texttt{COSINUS} [$1$] & \num{3.63e-14} & \num{6.75e-14} & \num{5.21e-14} & \num{3.63e-14} & \num{6.75e-14} & \num{5.21e-14} \\
Matrix, one group, \texttt{CUBIC} [$1$] & \num{4.30e-14} & \num{2.69e-14} & \num{5.30e-14} & \num{4.30e-14} & \num{2.69e-14} & \num{5.30e-14} \\
\noalign{\vskip 3pt\hrule height .6pt}
\end{tabular}%
}
\end{table}
\clearpage

\begin{table}[tbp]
\centering
\small
\setlength{\tabcolsep}{3pt}
\renewcommand{\arraystretch}{1.10}
\textbf{Table A1 (continued)}\par\medskip
\resizebox{\linewidth}{!}{%
\begin{tabular}{@{}>{\raggedright\arraybackslash}p{0.32\textwidth}rrrrrr@{}}
\noalign{\hrule height .6pt\vskip 3pt}
\textbf{Scalar quantity / option (unit)} & \multicolumn{3}{c}{\textbf{CPU}} & \multicolumn{3}{c}{\textbf{GPU}} \\

 & Value & Gradient & Deformation & Value & Gradient & Deformation \\
\noalign{\vskip 2pt\hrule height .35pt\vskip 2pt}
Matrix, two groups, $M_{11}$ [$1$] & \num{5.37e-14} & \num{6.62e-14} & \num{1.28e-13} & \num{5.37e-14} & \num{6.62e-14} & \num{1.28e-13} \\
Matrix, two groups, $M_{12}$ [$1$] & \num{8.82e-14} & \num{1.30e-13} & \num{1.43e-13} & \num{8.82e-14} & \num{1.30e-13} & \num{1.43e-13} \\
Matrix, two groups, $M_{21}$ [$1$] & \(0\) & \(0\) & \(0\) & \(0\) & \(0\) & \(0\) \\
Matrix, two groups, $M_{22}$ [$1$] & \num{7.27e-14} & \num{1.33e-13} & \num{2.17e-13} & \num{7.27e-14} & \num{1.33e-13} & \num{2.17e-13} \\
Matrix, two groups, \texttt{EXP} [$1$] & \num{2.88e-14} & \num{2.13e-14} & \num{2.58e-14} & \num{2.89e-14} & \num{2.13e-14} & \num{2.59e-14} \\
Matrix, two groups, \texttt{GAUSSIAN} [$1$] & \num{5.02e-14} & \num{2.35e-14} & \num{4.33e-14} & \num{5.02e-14} & \num{2.35e-14} & \num{4.35e-14} \\
Matrix, two groups, \texttt{TANH} [$1$] & \num{3.02e-14} & \num{2.97e-14} & \num{2.92e-14} & \num{3.03e-14} & \num{2.98e-14} & \num{2.93e-14} \\
Matrix, two groups, \texttt{SMAP} [$1$] & \num{8.09e-14} & \num{6.44e-14} & \num{7.46e-14} & \num{8.08e-14} & \num{6.43e-14} & \num{7.47e-14} \\
Matrix, two groups, \texttt{COSINUS} [$1$] & \num{1.68e-14} & \num{1.37e-13} & \num{1.22e-13} & \num{1.68e-14} & \num{1.37e-13} & \num{1.22e-13} \\
Matrix, two groups, \texttt{CUBIC} [$1$] & \num{5.36e-14} & \num{6.58e-14} & \num{1.55e-13} & \num{5.36e-14} & \num{6.58e-14} & \num{1.55e-13} \\
\noalign{\vskip 2pt\hrule height .35pt\vskip 2pt}
RMSD, translation [$\angstrom$] & \num{3.11e-14} & \num{8.55e-15} & \num{4.57e-14} & \num{3.11e-14} & \num{8.63e-15} & \num{4.57e-14} \\
RMSD, rotation [$\angstrom$] & \num{5.24e-14} & \num{5.71e-14} & \num{7.72e-14} & \num{4.97e-14} & \num{5.55e-14} & \num{7.62e-14} \\
Squared RMSD, translation [$\angstrom^2$] & \num{2.04e-13} & \num{4.66e-14} & \num{3.38e-13} & \num{2.04e-13} & \num{4.66e-14} & \num{3.38e-13} \\
Squared RMSD, rotation [$\angstrom^2$] & \num{2.45e-13} & \num{2.78e-13} & \num{3.89e-13} & \num{2.32e-13} & \num{2.82e-13} & \num{3.88e-13} \\
RMSD, translation, equal weights [$\angstrom$] & \num{3.42e-14} & \num{7.32e-15} & \num{5.05e-14} & \num{3.46e-14} & \num{7.33e-15} & \num{5.10e-14} \\
RMSD, rotation, equal weights [$\angstrom$] & \num{2.82e-14} & \num{1.39e-14} & \num{3.76e-14} & \num{2.78e-14} & \num{1.41e-14} & \num{3.76e-14} \\
\noalign{\vskip 2pt\hrule height .35pt\vskip 2pt}
Path progress [$1$] & \num{5.02e-14} & \num{4.10e-14} & \num{9.05e-14} & \num{5.77e-14} & \num{4.07e-14} & \num{9.48e-14} \\
Path distance [$\angstrom^2$] & \num{9.28e-14} & \num{3.41e-14} & \num{1.20e-13} & \num{9.37e-14} & \num{3.46e-14} & \num{1.22e-13} \\
\noalign{\vskip 2pt\hrule height .35pt\vskip 2pt}
Angular tetrahedral order [$1$] & \num{2.41e-13} & \num{2.83e-13} & \num{2.53e-13} & \num{2.41e-13} & \num{2.82e-13} & \num{2.56e-13} \\
\noalign{\vskip 2pt\hrule height .35pt\vskip 2pt}
$Q_1$, \texttt{PLUMED} normalization [$1$] & \num{1.43e-14} & \num{3.66e-14} & \num{2.30e-14} & \num{1.43e-14} & \num{3.66e-14} & \num{2.30e-14} \\
$Q_1$, conventional normalization [$1$] & \num{2.92e-14} & \num{7.49e-14} & \num{4.69e-14} & \num{2.93e-14} & \num{7.50e-14} & \num{4.71e-14} \\
$Q_3$, \texttt{PLUMED} normalization [$1$] & \num{4.20e-14} & \num{7.14e-14} & \num{1.42e-13} & \num{4.25e-14} & \num{7.14e-14} & \num{1.42e-13} \\
$Q_3$, conventional normalization [$1$] & \num{5.63e-14} & \num{9.63e-14} & \num{1.90e-13} & \num{5.70e-14} & \num{9.57e-14} & \num{1.89e-13} \\
$Q_4$, \texttt{PLUMED} normalization [$1$] & \num{4.50e-14} & \num{2.22e-13} & \num{1.11e-13} & \num{4.49e-14} & \num{2.22e-13} & \num{1.11e-13} \\
$Q_4$, conventional normalization [$1$] & \num{5.32e-14} & \num{2.63e-13} & \num{1.31e-13} & \num{5.30e-14} & \num{2.63e-13} & \num{1.31e-13} \\
$Q_6$, \texttt{PLUMED} normalization [$1$] & \num{4.19e-14} & \num{3.92e-13} & \num{1.68e-13} & \num{4.29e-14} & \num{3.92e-13} & \num{1.68e-13} \\
$Q_6$, conventional normalization [$1$] & \num{4.12e-14} & \num{3.85e-13} & \num{1.65e-13} & \num{4.22e-14} & \num{3.85e-13} & \num{1.65e-13} \\
\noalign{\vskip 2pt\hrule height .35pt\vskip 2pt}
Rational switch, removable limit [$1$] & \num{2.22e-16} & \num{6.35e-11} & \num{1.08e-10} & \num{2.22e-16} & \num{6.35e-11} & \num{1.08e-10} \\
Rational switch, raw [$1$] & \num{5.37e-14} & \num{6.62e-14} & \num{1.28e-13} & \num{5.37e-14} & \num{6.62e-14} & \num{1.28e-13} \\
Rational switch, raw cutoff [$1$] & \num{3.78e-14} & \num{1.01e-13} & \num{1.45e-13} & \num{3.78e-14} & \num{1.01e-13} & \num{1.45e-13} \\
Rational switch, stretched cutoff [$1$] & \num{4.17e-14} & \num{4.17e-14} & \num{8.14e-14} & \num{4.17e-14} & \num{4.17e-14} & \num{8.14e-14} \\
\noalign{\vskip 2pt\hrule height .35pt\vskip 2pt}
\texttt{EXP} switching function [$1$] & \num{2.03e-14} & \num{1.87e-14} & \num{2.56e-14} & \num{2.03e-14} & \num{1.87e-14} & \num{2.56e-14} \\
\noalign{\vskip 2pt\hrule height .35pt\vskip 2pt}
\texttt{GAUSSIAN} switching function [$1$] & \num{4.24e-14} & \num{2.21e-14} & \num{5.01e-14} & \num{4.24e-14} & \num{2.21e-14} & \num{5.01e-14} \\
\noalign{\vskip 2pt\hrule height .35pt\vskip 2pt}
\texttt{TANH} switching function [$1$] & \num{2.28e-14} & \num{2.27e-14} & \num{2.90e-14} & \num{2.27e-14} & \num{2.27e-14} & \num{2.90e-14} \\
\noalign{\vskip 2pt\hrule height .35pt\vskip 2pt}
\texttt{SMAP} switching function [$1$] & \num{4.20e-14} & \num{5.88e-14} & \num{1.08e-13} & \num{4.20e-14} & \num{5.88e-14} & \num{1.08e-13} \\
\noalign{\vskip 2pt\hrule height .35pt\vskip 2pt}
\texttt{COSINUS} switching function [$1$] & \num{3.63e-14} & \num{6.75e-14} & \num{5.22e-14} & \num{3.63e-14} & \num{6.75e-14} & \num{5.22e-14} \\
\noalign{\vskip 2pt\hrule height .35pt\vskip 2pt}
\texttt{CUBIC} switching function [$1$] & \num{4.30e-14} & \num{2.69e-14} & \num{5.30e-14} & \num{4.30e-14} & \num{2.69e-14} & \num{5.30e-14} \\
\noalign{\vskip 3pt\hrule height .6pt}
\end{tabular}%
}
\end{table}
\clearpage

\begin{table}[tbp]
\centering
\small
\setlength{\tabcolsep}{3pt}
\renewcommand{\arraystretch}{1.10}
\textbf{Table A1 (continued)}\par\medskip
\resizebox{\linewidth}{!}{%
\begin{tabular}{@{}>{\raggedright\arraybackslash}p{0.32\textwidth}rrrrrr@{}}
\noalign{\hrule height .6pt\vskip 3pt}
\textbf{Scalar quantity / option (unit)} & \multicolumn{3}{c}{\textbf{CPU}} & \multicolumn{3}{c}{\textbf{GPU}} \\

 & Value & Gradient & Deformation & Value & Gradient & Deformation \\
\noalign{\vskip 2pt\hrule height .35pt\vskip 2pt}
Constant scalar [$1$] & \(0\) & \(0\) & \(0\) & \(0\) & \(0\) & \(0\) \\
Constant vector, component 1 [$1$] & \(0\) & \(0\) & \(0\) & \(0\) & \(0\) & \(0\) \\
Constant vector, component 2 [$1$] & \(0\) & \(0\) & \(0\) & \(0\) & \(0\) & \(0\) \\
Constant vector, component 3 [$1$] & \(0\) & \(0\) & \(0\) & \(0\) & \(0\) & \(0\) \\
\noalign{\vskip 2pt\hrule height .35pt\vskip 2pt}
\texttt{COMBINE}, linear [$\angstrom$] & \num{1.45e-13} & \num{7.03e-14} & \num{2.10e-13} & \num{1.46e-13} & \num{7.04e-14} & \num{2.10e-13} \\
\texttt{COMBINE}, mixed powers [numeric] & \num{2.36e-12} & \num{8.90e-13} & \num{4.01e-12} & \num{2.37e-12} & \num{8.90e-13} & \num{3.99e-12} \\
\texttt{COMBINE}, shifted powers [numeric] & \num{2.23e-12} & \num{8.76e-13} & \num{3.87e-12} & \num{2.23e-12} & \num{8.76e-13} & \num{3.87e-12} \\
\texttt{COMBINE}, normalized [numeric] & \num{7.80e-13} & \num{3.63e-13} & \num{1.63e-12} & \num{7.78e-13} & \num{3.63e-13} & \num{1.64e-12} \\
\noalign{\vskip 2pt\hrule height .35pt\vskip 2pt}
\texttt{CUSTOM}, arithmetic [numeric] & \num{8.53e-13} & \num{2.59e-13} & \num{1.35e-12} & \num{8.53e-13} & \num{2.59e-13} & \num{1.35e-12} \\
\texttt{CUSTOM}, $\pi$ [$\angstrom$] & \num{1.07e-13} & \num{4.96e-14} & \num{1.25e-13} & \num{1.07e-13} & \num{4.96e-14} & \num{1.25e-13} \\
\texttt{CUSTOM}, sin [$1$] & \num{7.48e-14} & \num{5.70e-14} & \num{1.52e-13} & \num{7.44e-14} & \num{5.71e-14} & \num{1.52e-13} \\
\texttt{CUSTOM}, cos [$1$] & \num{7.95e-14} & \num{6.96e-14} & \num{2.40e-13} & \num{7.96e-14} & \num{6.93e-14} & \num{2.40e-13} \\
\texttt{CUSTOM}, tan [$1$] & \num{1.08e-14} & \num{4.83e-15} & \num{1.39e-14} & \num{1.08e-14} & \num{4.83e-15} & \num{1.39e-14} \\
\texttt{CUSTOM}, exp [$1$] & \num{1.29e-14} & \num{5.37e-15} & \num{1.79e-14} & \num{1.29e-14} & \num{5.37e-15} & \num{1.80e-14} \\
\texttt{CUSTOM}, log [$1$] & \num{3.15e-14} & \num{2.12e-14} & \num{3.53e-14} & \num{3.15e-14} & \num{2.12e-14} & \num{3.53e-14} \\
\texttt{CUSTOM}, sqrt [numeric] & \num{2.84e-14} & \num{1.58e-14} & \num{3.15e-14} & \num{2.84e-14} & \num{1.58e-14} & \num{3.16e-14} \\
\texttt{CUSTOM}, abs [$\angstrom$] & \num{1.74e-13} & \num{5.97e-14} & \num{1.80e-13} & \num{1.74e-13} & \num{5.95e-14} & \num{1.80e-13} \\
\texttt{CUSTOM}, tanh [$1$] & \num{1.34e-14} & \num{1.79e-14} & \num{1.80e-14} & \num{1.37e-14} & \num{1.81e-14} & \num{1.82e-14} \\
\texttt{CUSTOM}, min [$\angstrom$] & \num{1.04e-13} & \num{4.75e-14} & \num{1.02e-13} & \num{1.04e-13} & \num{4.75e-14} & \num{1.02e-13} \\
\texttt{CUSTOM}, max [$\angstrom$] & \num{1.19e-13} & \num{4.42e-14} & \num{1.19e-13} & \num{1.19e-13} & \num{4.43e-14} & \num{1.19e-13} \\
\noalign{\vskip 2pt\hrule height .35pt\vskip 2pt}
Harmonic distance bias [$\kjmol$] & \num{4.37e-13} & \num{1.91e-13} & \num{9.66e-13} & \num{4.37e-13} & \num{1.91e-13} & \num{9.66e-13} \\
Harmonic torsion bias [$\kjmol$] & \num{6.86e-13} & \num{5.47e-13} & \num{4.69e-13} & \num{6.86e-13} & \num{5.46e-13} & \num{4.69e-13} \\
\noalign{\vskip 2pt\hrule height .35pt\vskip 2pt}
Upper wall bias [$\kjmol$] & \num{1.88e-12} & \num{7.92e-13} & \num{4.08e-12} & \num{1.88e-12} & \num{7.89e-13} & \num{4.07e-12} \\
Lower wall bias [$\kjmol$] & \num{1.46e-13} & \num{3.27e-13} & \num{2.21e-13} & \num{1.46e-13} & \num{3.27e-13} & \num{2.21e-13} \\
\noalign{\vskip 3pt\hrule height .6pt}
\end{tabular}%
}
\end{table}
\clearpage

\clearpage
\subsection{Validation of umbrella sampling and metadynamics}
\label{app:bias-validation}
\noindent We test umbrella sampling (US) and metadynamics (MTD) for one particle of mass 12 atomic mass units in the one-dimensional potential $U(x)=4\,\kjmol[(x/\angstrom)^2-1]^2$, where $x$ is the particle's Cartesian position along the sampled direction.
The exact free energy is $F(x)=U(x)+C$, where $C$ sets the arbitrary energy zero.
It has equal minima at $x=\pm1\,\angstrom$ and a barrier of $4\,\kjmol$.
All four sampling variants use \texttt{JAX-MD} to integrate Langevin dynamics at 300~K with a 1~fs timestep and friction $5\,\mathrm{ps}^{-1}$~\cite{schoenholz_jax_2020}.
Standard and replica exchange US (REUS) each use 19 windows spaced by $0.200\,\angstrom$ from $-1.8$ to $+1.8\,\angstrom$, with stiffness $50\,\kjmol\,\angstrom^{-2}$, 20~ps of equilibration, and 1~ns of production per window.
The exchange variant attempts swaps between alternating neighboring pairs every 20~fs and refreshes the forces after accepted swaps; the acceptance fraction is 0.589.
Three independently seeded window sets are batched together, and the free energy profile for each set is reconstructed separately using the multistate Bennett acceptance ratio.

\noindent We compare single walker well-tempered MTD for 4~ns with four walkers sharing one bias for 1~ns each, giving the same aggregate sampling time of 4~ns per independent repeat.
Three independently seeded repeats use 20~ps of unbiased equilibration, hill width $0.200\,\angstrom$, initial height $0.050\,\kjmol$ per hill, bias factor 5, and deposition every 20~fs per walker.
Thus, the per-hill parameters and nominal hill counts match at equal aggregate sampling, while the actual well-tempered heights depend on the evolving bias.
The bias grid spans $[-3,3]\,\angstrom$ with 1,200 intervals.
MTD profiles use the recorded bias and $c(t)$ weights~\cite{tiwary_timeindependent_2015} after discarding the initial 10\% of sampled configurations; all deposited hills remain in the bias.

\begin{figure}[tbp]
\centering
\includegraphics[width=0.65\linewidth]{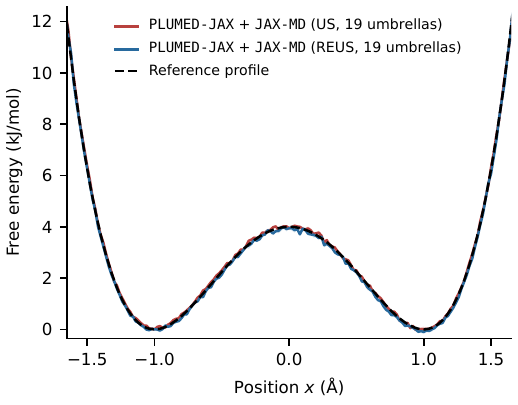}
\caption{Double-well free energy at 300~K from standard and replica exchange umbrella sampling, each using 19 windows of 1~ns.
The one-particle potential is $U(x)=4\,\kjmol[(x/\angstrom)^2-1]^2$, where $x$ is the particle position; profiles are reconstructed using the multistate Bennett acceptance ratio.
Curves are means of three independent repeat profiles, using $0.0200\,\angstrom$ bins and a common energy zero at $x=-1\,\angstrom$.
Shading is one standard error across repeats; the dashed curve is the exact bin-integrated free energy.}
\label{fig:double-well-umbrella}
\end{figure}

\begin{figure}[tbp]
\centering
\includegraphics[width=0.65\linewidth]{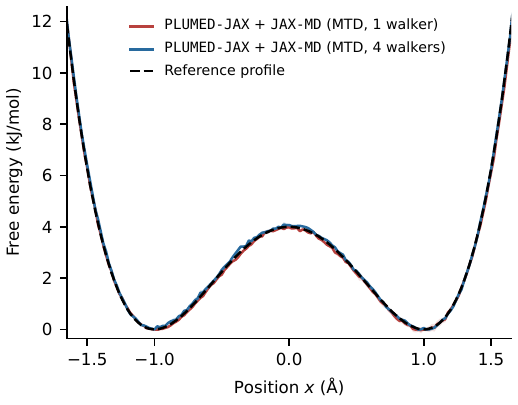}
\caption{Double-well free energy at 300~K from single walker well-tempered metadynamics for 4~ns and four walkers sharing a bias for 1~ns each.
The one-particle potential is $U(x)=4\,\kjmol[(x/\angstrom)^2-1]^2$, where $x$ is the particle position; histograms are reweighted with the metadynamics bias and its time-dependent normalization $c(t)$.
Curves are means of three independent repeat profiles, using $0.0200\,\angstrom$ bins and a common energy zero at $x=-1\,\angstrom$.
Shading is one standard error across independent biases; the dashed curve is the exact bin-integrated free energy.
Reweighted histograms discard the first 10\% of samples while keeping all deposited hills in the bias.}
\label{fig:double-well-MTD}
\end{figure}

\noindent Figures~\ref{fig:double-well-umbrella} and \ref{fig:double-well-MTD} compare mean profiles from the three independent repeats, using $0.0200\,\angstrom$ bins and a common energy zero at $x=-1\,\angstrom$.
The root-mean-square errors relative to the exact bin-integrated profile, over coordinates with $F(x)-\min F\leq5k_{\mathrm B}T$, are $0.0627$ and $0.0813\,\kjmol$ for standard and replica exchange umbrellas, and $0.0658$ and $0.0702\,\kjmol$ for single walker and multiple walker MTD, respectively.

\clearpage
\section{Parallel execution of lithium umbrellas}
\label{app:window-scaling}
\begin{figure}[tbp]
\centering
\includegraphics[width=0.9\linewidth]{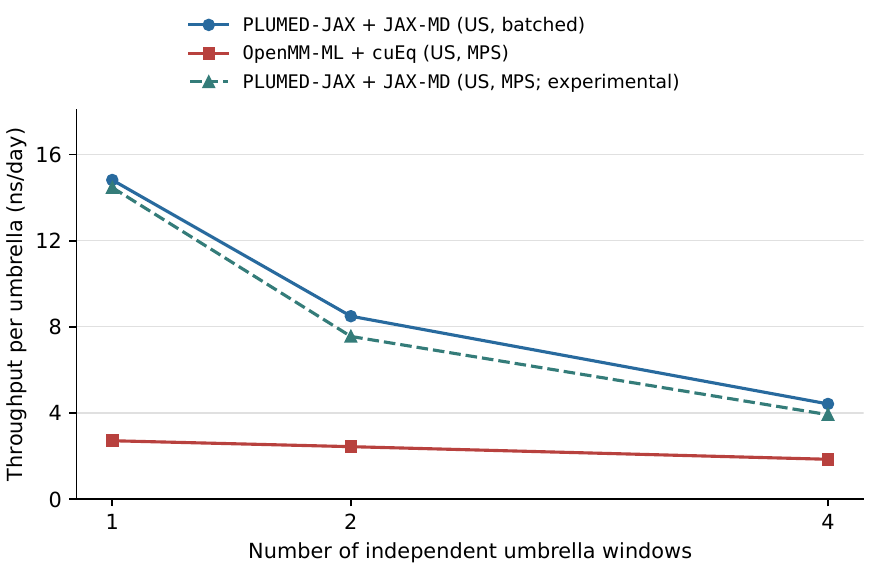}
\caption{Throughput per lithium umbrella on one RTX A6000, obtained by dividing aggregate throughput by the number of windows.
All cases use the \texttt{MACE-MP-0-medium} potential for 601 atoms and independent harmonic umbrellas at 450~K, without replica exchange.}
\label{fig:lithium-window-scaling}
\end{figure}

\noindent Figure~\ref{fig:lithium-window-scaling} compares batched \texttt{JAX-MD} using \texttt{jax.vmap} with separate \texttt{OpenMM-ML} or \texttt{JAX-MD} processes under \texttt{NVIDIA Multi-Process Service} (\texttt{MPS}), for one, two, or four independent lithium umbrellas on one RTX A6000.
The single-window \texttt{JAX} control also uses \texttt{jax.vmap}.
The \texttt{JAX}+\texttt{MPS} calculation is used as a diagnostic of GPU usage; its driver is not included in the repository.
No calculation in this comparison uses replica exchange.

\noindent With batched \texttt{JAX}, the rate per umbrella decreases from 14.8 to 4.42~ns/day as the number of windows increases from one to four; aggregate throughput increases by a factor of 1.19.
For \texttt{OpenMM}+\texttt{MPS}, the corresponding rates are 2.71 and 1.85~ns/day.
\texttt{JAX}+\texttt{MPS} gives 3.92~ns/day per umbrella at four windows, below the batched \texttt{JAX} result.

\noindent All cases use the same \texttt{MACE-MP-0-medium} checkpoint, 32-bit model inference, 64-bit bias and integration, a 1~fs timestep, a 450~K Langevin thermostat with friction $1\,\mathrm{ps}^{-1}$, and umbrella stiffness $89.8\,\kjmol\,\angstrom^{-2}$, with \texttt{cuEquivariance} acceleration.
Window centers are spaced by $0.250\,\angstrom$, symmetrically around the CV value in the common starting structure.
Each process has four CPU cores available, with \texttt{Torch} and numerical library thread counts set to one.

\noindent Each window is timed over 100 steps after warm-up, excluding trajectory output.
Aggregate throughput is the sum of simulated time across windows divided by the elapsed time from the first worker's start to the last worker's finish.

\end{document}